\documentclass[aps,prr,twocolumn,superscriptaddress,longbibliography]{revtex4-2}

\usepackage{mathrsfs,amssymb,graphics,subfigure,threeparttable}
\usepackage{grap hicx}
\usepackage{amssymb}
\usepackage{enumerate}
\usepackage{amsmath}
\usepackage{fancyhdr}
\usepackage{dcolumn}
\usepackage{bm}
\usepackage[squaren]{SIunits}
\usepackage{hyperref}

\usepackage{xcolor}

\begin{document}


\title{Highly spin-polarized multi-GeV electron beams generated by single-species plasma photocathodes}


\author{Zan Nie}
\email[]{znie@ucla.edu}
\affiliation{Department of Electrical and Computer Engineering, University of California Los Angeles, Los Angeles, California 90095, USA}
\author{Fei Li}
\email[]{lifei11@ucla.edu}
\affiliation{Department of Electrical and Computer Engineering, University of California Los Angeles, Los Angeles, California 90095, USA}

\author{Felipe Morales}
\author{Serguei Patchkovskii}
\author{Olga Smirnova}
\affiliation{Max Born Institute, Max-Born-Str. 2A, D-12489 Berlin, Germany}

\author{Weiming An}
\affiliation{Department of Astronomy, Beijing Normal University, Beijing 100875, China}

\author{Chaojie Zhang}
\author{Yipeng Wu}
\author{Noa Nambu}
\author{Daniel Matteo}
\author{Kenneth A. Marsh}
\affiliation{Department of Electrical and Computer Engineering, University of California Los Angeles, Los Angeles, California 90095, USA}

\author{Frank Tsung}
\affiliation{Department of Physics and Astronomy, University of California Los Angeles, Los Angeles, California 90095, USA}
\author{Warren B. Mori}
\affiliation{Department of Electrical and Computer Engineering, University of California Los Angeles, Los Angeles, California 90095, USA}
\affiliation{Department of Physics and Astronomy, University of California Los Angeles, Los Angeles, California 90095, USA}

\author{Chan Joshi}
\email[]{cjoshi@ucla.edu}
\affiliation{Department of Electrical and Computer Engineering, University of California Los Angeles, Los Angeles, California 90095, USA}

\date{\today}

\begin{abstract}

High-gradient and high-efficiency acceleration in plasma-based accelerators has been demonstrated, showing its potential as the building block for a future collider operating at the energy frontier of particle physics. However, generating and accelerating the required spin-polarized beams in such a collider using plasma-based accelerators has been a long-standing challenge. Here we show that the passage of a highly relativistic, high-current electron beam through a single-species (ytterbium) vapor excites a nonlinear plasma wake by primarily ionizing the two outer $6s$ electrons. Further photoionization of the resultant Yb$^{2+}$ ions by a circularly polarized laser injects the $4f^{14}$ electrons into this wake generating a highly spin-polarized beam. Combining time-dependent Schr\"{o}dinger equation simulations with particle-in-cell simulations, we show that a sub-femtosecond, high-current (4\,kA) electron beam with up to 56\% net spin polarization can be generated and accelerated to 15\,GeV in just 41\,cm. This relatively simple scheme solves the perplexing problem of producing spin-polarized relativistic electrons in plasma-based accelerators.

\end{abstract}


\maketitle

\section{Introduction}
High-brightness relativistic spin-polarized beams play indispensable roles in high energy physics, such as in high-energy lepton colliders \cite{barish2013international} and parity violation experiments \cite{Souder2016,Adhikari2021}. However, conventional radio-frequency based accelerators that can reach the necessary energies at the energy frontier of collider-based particle physics will become gargantuan and prohibitively expensive. By demonstrating orders of magnitude higher accelerating gradient and a high energy extraction efficiency,   plasma-based accelerator (PBA) offers a paradigm changing alternative that promises to shrink the size and cost of future high-energy colliders \cite{Joshi2021}. However, a practical scheme for generating and accelerating spin-polarized leptons in PBAs is still lacking. Conventional methods of generating spin-polarized electrons include self-polarization via the Sokolov-Ternov effect \cite{sokolov1964polarization}, photoionization of alkali atoms \cite{Long1965}, Fano effect \cite{fano1969spin}, Mott scattering \cite{tolhoek1956electron}, and photoemission from a Gallium Arsenide (GaAs) cathode \cite{pierce1976photoemission}. However, none of these conventional methods can generate ultra-short (few microns long), high-current, and precisely synchronized (fs) spin-polarized electron beams needed for injection into PBAs. 

Recently, a down-ramp injection scheme using hydrogen halide gas for producing spin-polarized electrons in PBAs has been proposed \cite{wen2019polarized,wu2019polarized,wu2019polarized1}. However, in this two-step scheme, multiple laser beams are needed to first produce a plasma with polarized electrons. Also, the pre-polarized electrons can be easily depolarized in the down-ramp injection process, which limits both the accelerating gradient and charge of the injected electrons. In another proposal, a one-step scheme based on spin-polarized ionization injection \cite{Nie2021,*nie2021erratum} suitable for a beam-driven plasma wakefield accelerator (PWFA) that uses a mixture of xenon (Xe) and lithium (Li) was proposed. Unfortunately, this multi-species scheme is difficult to realize in practice and in any case the spin polarization is limited to $\sim 30\%$. 

Here we show a simpler scheme that can achieve a higher degree of spin polarization at the same time. We propose to use a single atomic species – ytterbium (Yb) to act as a plasma photocathode \cite{kaushal2018looking,kaushal2018looking1}, both for wake formation and ionization injection of spin-polarized electrons in a PWFA. Moreover, first experimental demonstration of high-quality, highly spin-polarized electron beams is now possible through this scheme using state-of-the-art high-energy beam facilities such as the Facility for Advanced Accelerator Tests II (FACET-II) \cite{Joshi2018} at SLAC National Accelerator Laboratory.

\section{Proposed scheme}
The proposed scheme takes advantage of electron spin polarization resulting from the sensitivity of strong-field ionization of atoms or ions by intense circularly polarized laser fields to the orbital angular momentum of initial bound orbitals \cite{barth2013spin,hartung2016electron,barth2013nonadiabatic,herath2012strong,eckart2018ultrafast,trabert2018spin}. Electrons removed from $p$, $d$ or $f$ orbitals can form short spin-polarized bunches, while electrons removed from $s$ orbitals are not spin-polarized. We consider Yb atom with the electron configuration of [Xe]\,$4f^{14} 6s^2$, with the ionization potentials of the $6s^2$, $6s^1$, and $4f^{14}$ electrons being 6.25, 12.18, and 25.05\,eV, respectively. Capitalizing on the significant difference in the ionization potentials for $s$ and $f$ electrons, we adjust the driving electron beam current such that its transverse electric field liberates primarily the two outer $6s$ electrons of Yb, leaving $4f^{14}$ electrons largely bound. These two liberated but unpolarized $6s$ electrons and Yb$^{2+}$ ions form the plasma. If the driving electron bunch is ultrarelativistic ($\gamma \gg1$) and sufficiently dense ($n_b>n_p$, $k_p \sigma_{r,z}<1$), then the plasma electrons are blown transversely by the collective Coulomb repulsion of the beam electrons to resonantly excite a bubble-like wake cavity \cite{lu2006nonlinear}, containing mostly the Yb$^{2+}$ (i.e. Yb III) ions. Here $n_b$ and $n_p$ are beam and plasma densities, $k_p$, $\sigma_r$ and $\sigma_z$ are the wavenumber of the linear wake, the r.m.s. bunch radius and bunch length respectively. A 400\,nm circularly polarized (CP) laser pulse, following the driving electron beam at a specific time delay arrives at the position in the bubble where the on-axis longitudinal electric field of the wake is zero, liberates the outermost $4f^{14}$ electrons of the Yb$^{2+}$ ions. These ionized spin-polarized electrons are then trapped near the rear of the first bucket of the wake and accelerated by the wake to multi-GeV energy.

\section{TDSE simulations and results}
Theoretical advances and recent experiments on ionization of atoms in strong laser fields indicate that spin-polarized electrons can be produced by strong-field ionization because the ionization probability in CP fields depends on the sense of electron rotation (i.e. the magnetic quantum number $m_l$) in the initial state \cite{barth2013spin,hartung2016electron,barth2013nonadiabatic,herath2012strong,eckart2018ultrafast,trabert2018spin}. This mechanism can operate for a broad range of laser frequencies and intensities. In our scheme (ionization of the $f$-orbital electrons of the Yb$^{2+}$ ions using a 400\,nm CP laser), we operate in the nonadiabatic tunneling regime, where ionization of counter-rotating electrons is dominant over that of co-rotating electrons. Using Yb$^{2+}$ yields a substantially higher degree of net spin polarization , because $f$-orbital electrons in Yb$^{2+}$ have a higher angular momentum and hence stronger spin-orbit coupling \cite{kaushal2018looking,kaushal2018looking1} than lower orbital ($p$ or $d$) electrons in noble gases. 

\begin{table}[b]
\caption{\label{tab:potential}%
Effective-potential fit parameters
}
\begin{ruledtabular}
\begin{tabular}{cccccc}
\textrm{J}&
\textrm{$r_{min}$}&
\textrm{a}&
\textrm{b}&
\textrm{c}&
\textrm{d}\\
\colrule
7/2 & 0.159936 & 45.2479500 & 1.0 & 0.25477 & -9.9480672 \\
5/2 & 0.161705 & 47.9445593 & 1.0 & 0.25019 & -10.3512619\\
\end{tabular}
\end{ruledtabular}
\end{table}

\begin{figure*}[tp]
\includegraphics[width=0.85\textwidth]{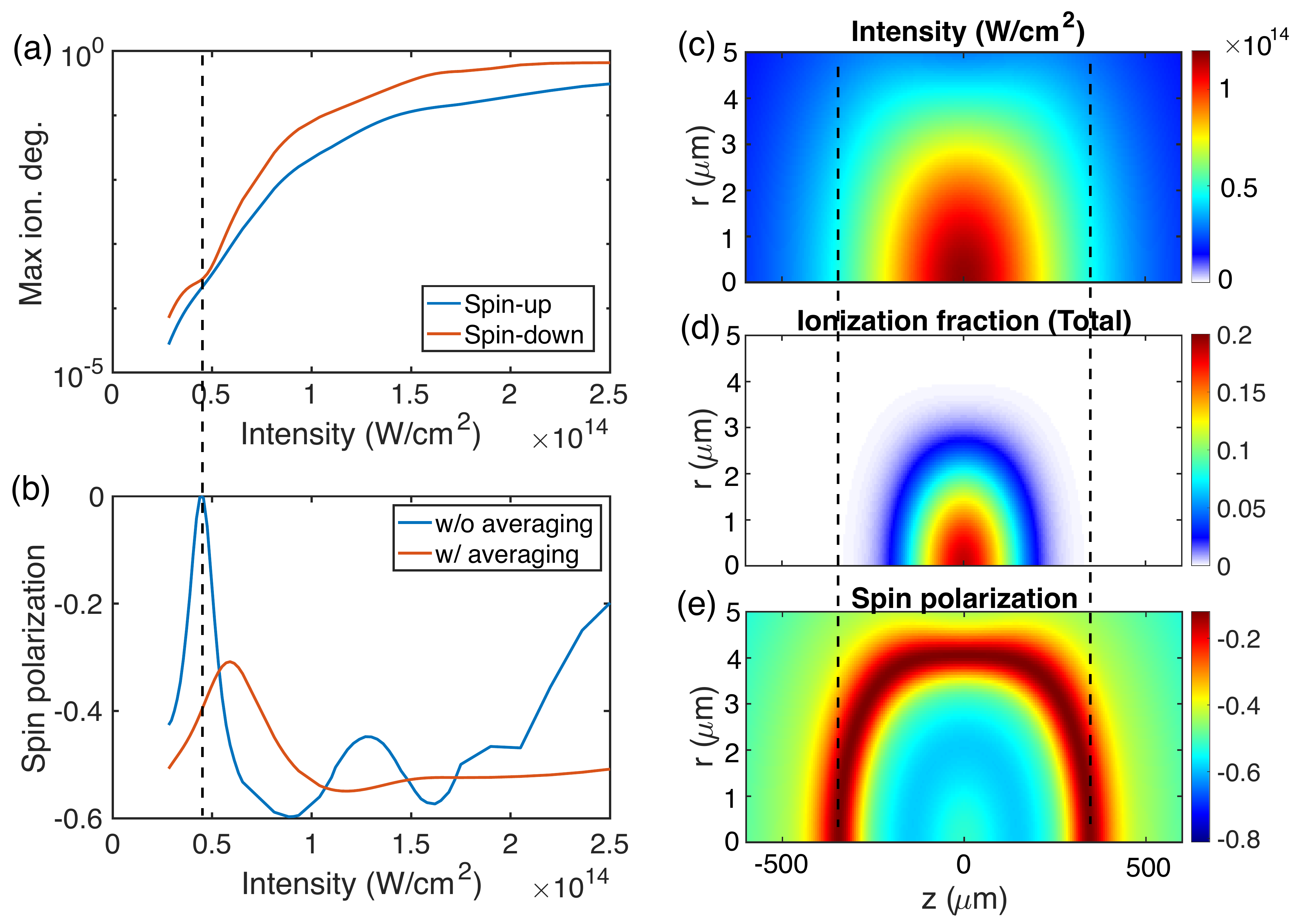}
\caption{\label{fig1} 
TDSE simulation results of ionization of the $4f^{14}$ electron of the Yb$^{2+}$ ion. (a), Maximum ionization degree of spin-up and spin-down electrons as a function of laser peak intensity of a 400\,nm, 60\,fs (FWHM), right-handed CP laser that ionizes spin-down electrons preferentially. (b), Spin polarization (on-axis) as a function of peak laser intensity without (blue: 1D) and with (red: 3D) focal-volume averaging in the region of $r<w_0$ and $|z| <$ Rayleigh length. The dashed line in (a) and (b) corresponds to the dip of spin polarization at $\sim 0.5\times 10^{14}$\,W/cm$^2$. (c-e), The laser intensity distribution, total ionization fraction distribution and spin polarization distribution in the $r$-$z$ plane for a bi-Gaussian 60\,fs (FWHM), $w_0$ = 6.0\,$\mu$m, 400 nm laser with a peak intensity of $1.18\times 10^{14}$\,W/cm$^2$.  The dashed line in (c) to (e) corresponds to the same on-axis laser intensity as the dashed line in (a) and (b).
}
\end{figure*}

We use the time-dependent Schr\"{o}dinger equation (TDSE) code SCID \cite{patchkovskii2016simple} for a range of laser intensities from $2.8\times 10^{13}$\,W/cm$^2$ to $2.5\times 10^{14}$\,W/cm$^2$ for each ionization pathway, to calculate the corresponding spin-up and spin-down electron ionization rates and yields [Fig.\,\ref{fig1}(a)]. The potentials have been modelled to describe the ionization of Yb$^{2+}$, leaving the ion in a J=7/2 or a J=5/2 state. Energy of the $4f$ level in the J=7/2 channel was set at the experimental ionization potential of 25.053\,eV \cite{Lide2005}. The experimental spin-orbit splitting of 1.26637\,eV \cite{NIST_ASD} was used to set the $4f$ level in the J=5/2 channel. Experimental atomic excitation energies were taken from Ref.\,\cite{NIST_ASD} and supplemented with theoretical results from Ref.\,\cite{Safronova2009}. The effective potentials in the modified form \cite{Garvey1975} (atomic units) are:
\begin{align}\label{Potential}
u(r)&=\infty, \,(r\leq r_{min}) \\
u(r)&=-\frac{3}{r}+\frac{a}{r} \, \frac{1}{b+\frac{d}{c} [\exp(c r)-1]}, \,(r> r_{min})
\end{align}
The fit parameters for J=7/2 and J=5/2 states are summarized in Table\,\ref{tab:potential}. More details about TDSE simulations are shown in Appendix \ref{sec_TDSE}.

Here, we choose right-handed CP laser pulses in our TDSE simulations, and we define spin polarization as $(N_\uparrow-N_\downarrow)/(N_\uparrow+N_\downarrow )$, where $N_{\uparrow,\downarrow}$ represents the number of spin-up or spin-down (parallel or antiparallel to laser wave-vector $\vec{k}$) electrons. From TDSE simulations, the maximum ionization degrees of spin-up and spin-down electrons versus laser intensity are shown in Fig.\,\ref{fig1}(a). Accordingly, the net spin polarization after integration over the entire temporal and spatial intensity distribution of the laser pulse, all photoelectron energies, and all final ionic states is shown in Fig.\,\ref{fig1}(b). From Fig.\,\ref{fig1}(b), one can see that the net spin polarization can be higher than 50\% for a broad range of laser intensity (0.8-2.5)$\times 10^{14}$\,W/cm$^2$ after focal-volume averaging. Therefore, the injected electron charge can be increased/decreased by increasing/decreasing the laser intensity in this range while not appreciably changing the degree of spin polarization. The steep drop of spin polarization at $\sim 0.5\times 10^{14}$\,W/cm$^2$ is due to the Freeman resonances \cite{Freeman1987}. The $r$-$z$ spatial distributions of the total ionization fraction and spin polarization at the laser intensity of $1.18\times 10^{14}$\,W/cm$^2$ (which will be used in the later particle-in-cell (PIC) simulations) are shown in Fig.\,\ref{fig1}(c) and \ref{fig1}(d), respectively. From Fig.\,\ref{fig1}(c), one can see that the maximum ionization fraction of the $4f^{14}$ electrons of Yb is about 19\%. In Fig.\,\ref{fig1}(d), a dark red half ellipse representing low spin polarization at $\sim 0.5\times 10^{14}$\,W/cm$^2$ due to the Freeman resonance mentioned above is clearly observed, but the ionization yield of these electrons is low ($<10^{-3}$) so that the net spin polarization can still reach 56\% after focal-volume averaging.

\section{PIC simulations and results}
Next, we incorporate the spin-dependent ionization results into the wakefield acceleration simulations using OSIRIS \cite{fonseca2002osiris,fonseca2008one} and QPAD \cite{li2020quasi} codes. More details are shown in Appendix \ref{sec_TDSE_PIC} and \ref{sec_PIC}. We have implemented the spin precession module into both the OSIRIS and QPAD codes following the Thomas-Bargmann-Michel-Telegdi (T-BMT) equation \cite{bargmann1959precession}
\begin{align}\label{T-BMT}
d\bold{s}/dt=\bold{\Omega}\times \bold{s}
\end{align}
where $\bold{\Omega}=\frac{e}{m} (\frac{1}{\gamma} \bold{B}-\frac{1}{\gamma+1} \frac{\bold{v}}{c^2} \times \bold{E})+a_e\frac{e}{m} [\bold{B}-\frac{\gamma}{\gamma+1} \frac{\bold{v}}{c^2} (\bold{v}\cdot \bold{B})-\frac{\bold{v}}{c^2} \times \bold{E}]$ . Here, $\bold{E},\bold{B}$ are the electric and magnetic field, $\bold{v}$ is the electron velocity, $\gamma=\frac{1}{\sqrt{1-v^2/c^2} }$ is the relativistic factor, and $a_e\approx 1.16\times 10^{-3}$ is the anomalous magnetic moment of the electron. 

\begin{figure*}[tp]
\includegraphics[width=0.95\textwidth]{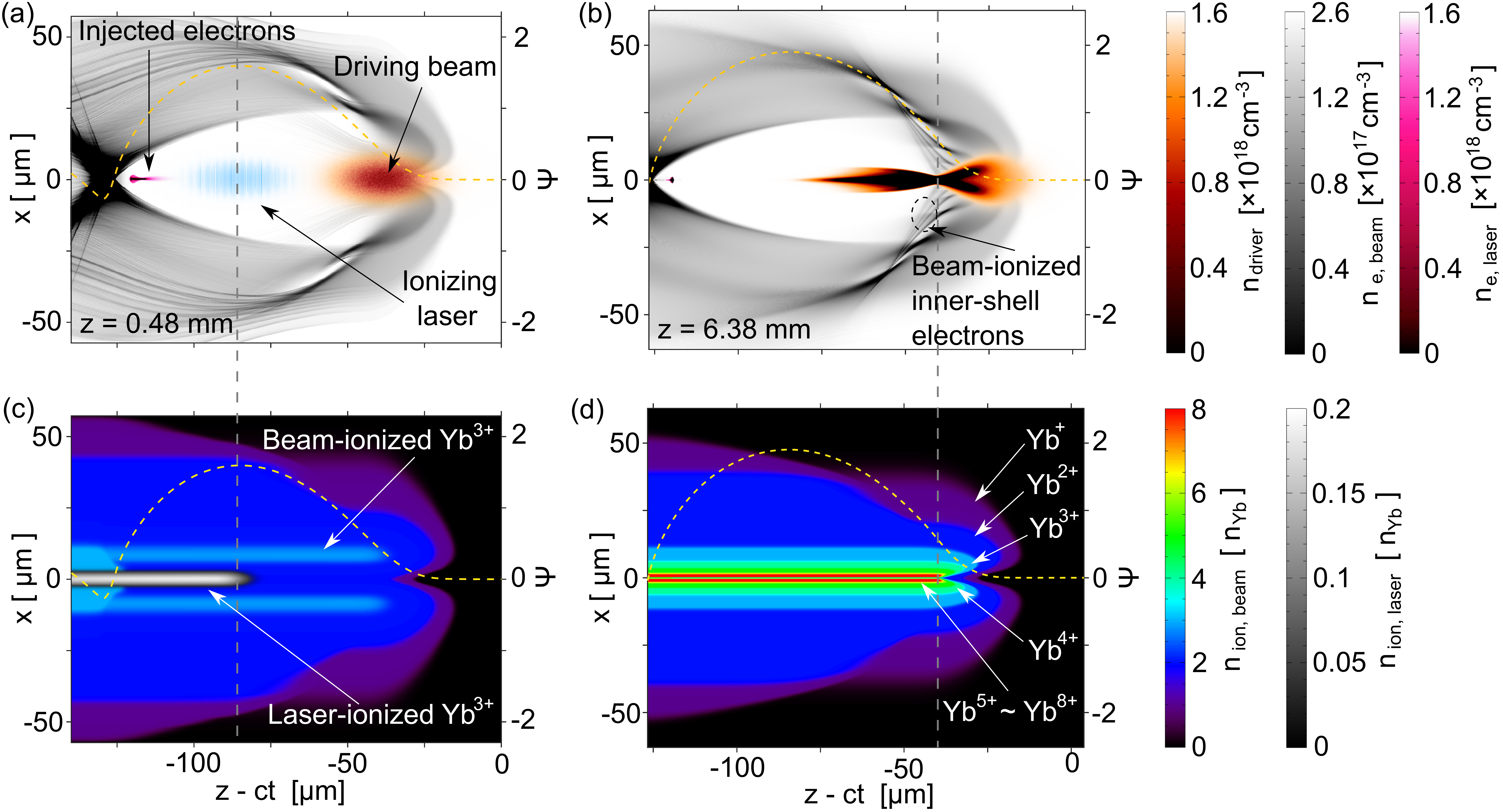}
\caption{\label{fig2} 
Snapshots of ionization injection and acceleration. (a,b), Two snapshots show the charge density distribution of driving electron beam (brown), beam-ionized Yb electrons (grey), and laser-ionized $4f^{14}$ electrons of Yb (purple) at (a) $z=0.48$\,mm (at around laser focus) and (b) $z=6.38$\,mm (driving beam pinched). (c,d), The Yb ion charge density distribution at the same moment of (a) and (b) respectively. The yellow dashed lines in (a-d) show the on-axis wake pseudo potential. (a) and (c) are from OSIRIS simulations. (b) and (d) are from QPAD simulations. (see Appendix).
}
\end{figure*}

\begin{figure*}[tp]
\includegraphics[width=0.85\textwidth]{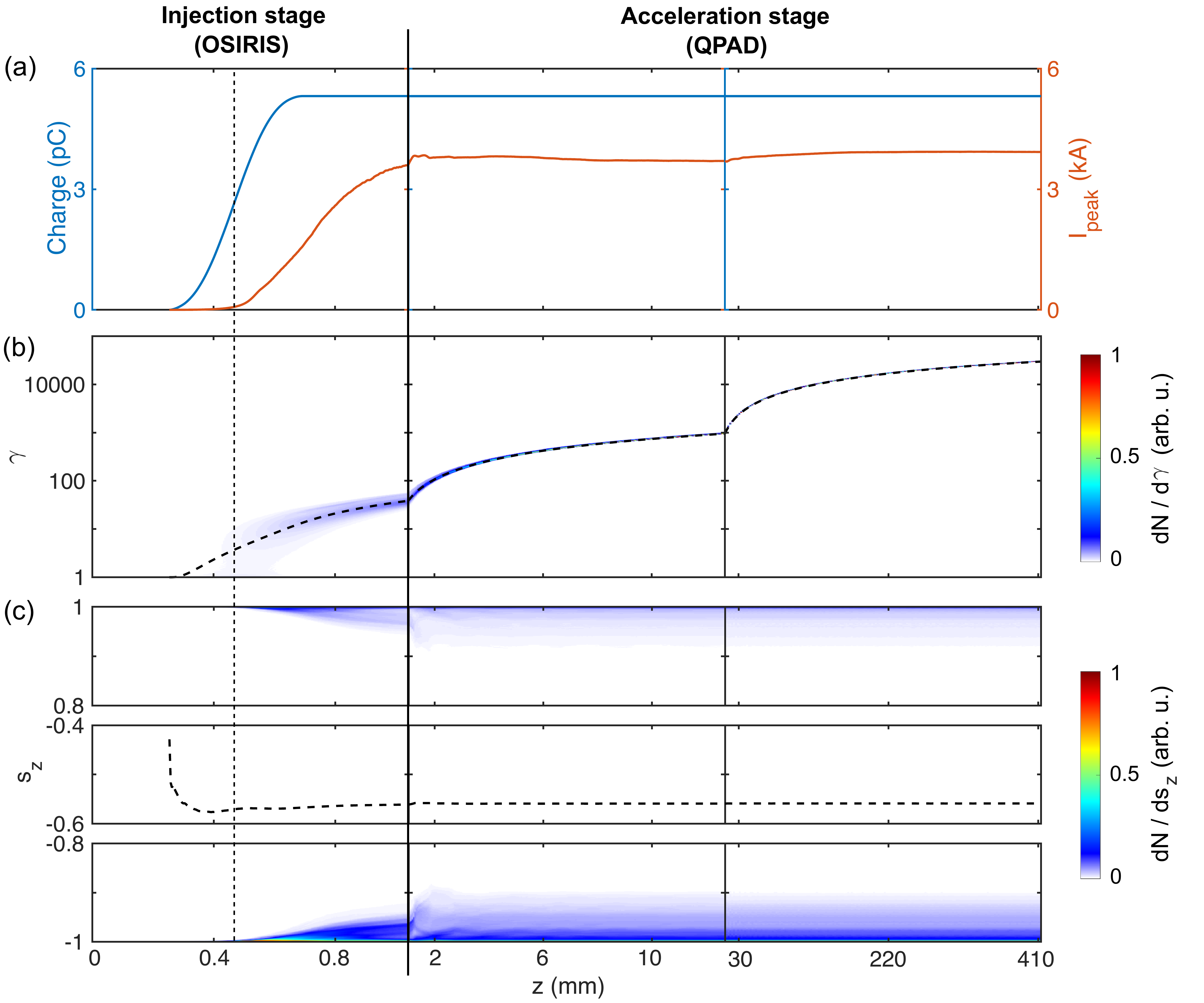}
\caption{\label{fig3} 
Evolution of injected electrons as a function of propagating distance. (a), Evolution of beam charge (left axis) and peak current (right axis). (b), Evolution of Lorentz factor $\gamma$. The dashed line presents mean energy $\langle \gamma\rangle$. (c), Evolution of spin vector in the z direction: $s_z$. The top box plots the $s_z$ distribution in the range of 0.8 and 1, corresponding to spin-up electron density. The central box plots $\langle s_z\rangle$ (net spin polarization) in the range of -0.6 and -0.4. The bottom box plots the $s_z$ distribution in the range of -1 and -0.8, corresponding to spin-down electron density. The long vertical dashed black line marks the focal position ($z=0.47$\,mm) of the ionization laser. The apparent discontinuities at the boundary of different stages are not real and appear only because the horizontal scale changes between different stages.
}
\end{figure*}

We carried out numerous PIC simulations where the parameters of the driving electron beam and ionizing laser were varied. Here, we present results for parameters which gave the best results but note that further optimization is possible. The driving electron beam has a Gaussian profile $n_b=\frac{N}{(2\pi)^{\frac{3}{2}}\sigma_r^2 \sigma_z} \exp(-\frac{r^2}{2\sigma_r^2}-\frac{\xi^2}{2\sigma_z^2})$ where $N=7.5\times10^9$ (1.2\,nC), $\sigma_r=6.4\,\mu$m and $\sigma_z=12.7\,\mu$m are the transverse and longitudinal beam sizes, respectively. The driving electron beam energy is 10\,GeV with normalized emittance of $\epsilon_n=16\,\mu$m. The transverse electric field of a relativistic electron beam with a Gaussian profile can be expressed as 
\begin{align}\label{Er}
E_r=\frac{1}{(2\pi)^{\frac{3}{2}}}  \frac{e}{\epsilon_0} \frac{N}{\sigma_r \sigma_z} \frac{1-\exp(-\frac{r^2}{2\sigma_r^2})}{r/\sigma_r} \exp(-\frac{z^2}{\sigma_z^2})
\end{align}
Such a transverse electric field vanishes at $r=0$ and has a maxima \cite{OConnell2006} $E_r^{max}=10.4 \frac{N}{10^{10}}  \frac{10}{\sigma_r [\mu m]}  \frac{50}{\sigma_z [\mu m]}$[GV/m] at $r\approx1.6\sigma_r$. When such a beam enters into very low-density neutral gases, it cannot ionize gas near the $r=0$ axis. However, in the simulations, the on-axis atoms are seen to be ionized by a combination of the transverse fields of the wake and the electric field of the driving beam (which can be enhanced due to self-focusing in the wake). In our simulations, the Yb gas has a uniform density of $n_{\text{Yb}}=5.22\times10^{16}$\,cm$^{-3}$ with an up-ramp length of 100\,$\mu$m (shorter than experimental values to save simulation time). In reality, if the ramp is about 10 cm long, the focus of the driving beam can be moved so that it does not pinch in the ramp \cite{Joshi2018}. By choosing such Yb gas density and driving beam parameters, we can get a large enough region around the axis ($r<7\,\mu$m) where the two $6s$ electrons of Yb are fully ionized [Fig.\,\ref{fig2}(c)]. It is this region that the follow-up ionizing laser can further ionize Yb$^{2+}$ to generate spin-polarized beams. At around $r\approx1.6\sigma_r\approx10\,\mu$m, a small fraction of the third electron ($4f^{14}$ electron) of Yb is also ionized but they are also blown out by the driving beam as shown in Fig.\,\ref{fig2}(c) and are not subsequently trapped. We note that at this early stage the ionization of the third electron is not caused by the pinching of the driving beam that occurs later from self-focusing. The driving electron beam blows out the first two and some third ionized electrons to create the wake cavity, leaving remaining Yb$^{2+}$ ions around the axis. The 400\,nm ionization laser with a pulse duration (FWHM) of 60\,fs and focal spot size of $w_0 = 6.0\,\mu$m, is delayed by 156\,fs (46.7\,$\mu$m) from the peak current position of the driving electron beam. This delay is chosen so that the laser is at the center of the wake bubble [Fig.\,\ref{fig2}(a)] and hence the best trapping condition is achieved ($\Delta\Psi=\Psi_f-\Psi_i\leq-1$, where $\Psi=e/(mc^2 )(\phi-A_z)$ is the normalized pseudopotential of the wake, and subscript ‘$i$’, ‘$f$’ indicates the value of the pseudopotential at the position of ionization and trapping, respectively) \cite{pak2010injection}. Here, $\phi$ is the electric potential and $A_z$ is the longitudinal component of the vector potential. The peak laser intensity is $1.18\times 10^{14}$\,W/cm$^2$ (the same intensity as in Fig.\,\ref{fig1}(c,d). 

As the unmatched (plasma ion focusing force $>$ diffraction caused by the beam emittance) driving electron beam propagates in the plasma, it is seen to pinch \cite{Xu2016}, leading to stronger local electric field and ionization of multiple $4f$ orbital electrons of Yb as shown in Fig.\,\ref{fig2}(b, d). Up to 8 electrons can be ionized by the pinched driving electron beam, but these extra ionized electrons are not trapped and accelerated by the wakefield (beam-induced ionization injection \cite{oz2007ionization,Vafaei-Najafabadi2019}) because the pulse length of our driving beam is short compared to the plasma cavity length; the location of beam-induced ionization is at the head of the wake where the difference between the initial and the final pseudo-potentials $\Delta\Psi$ is not sufficient to satisfy the trapping condition. To verify this, we have run another PIC simulation (using quasi-3D OSIRIS \cite{Lifschitz2009,Davidson2015} to save simulation time) with the same parameters but without the ionizing laser, and we found that no self-injection occurs in this case, i.e., no “dark current” exists. We have tried using driving electron beam with larger emittances or smaller spot sizes to minimize self-focusing. For larger emittances, the head of the beam diffracts preventing multi-GeV energy gain. For narrower spot sizes, the drive beam ionizes Yb$^{2+}$ closer to the axis such that the column of the Yb$^{2+}$ is too narrow and some of the unpolarized electrons get trapped. Therefore, the parameters we used here is a trade-off result between these two issues.

Evolution of injected beam parameters including charge, peak current, and spin vector distribution as a function of propagation distance in the plasma are shown in Fig.\,\ref{fig3}. Photoionized electrons with charge of 5.3\,pC (Fig.\,\ref{fig3}(a) left axis) are injected, trapped and accelerated to 15\,GeV [Fig.\,\ref{fig3}(b)] in 41\,cm until the driving beam is depleted of its energy. The pulse length of the injected bunch first increases to about $\sigma_{zi}$= 10\,$\mu$m, and then decreases to a final pulse length of only $\sigma_{zf}$= 0.2\,$\mu$m at the very back of the wake, corresponding to a sub-femtosecond bunch \cite{xu2016nanoscale}. The peak current is as high as 4\,kA (Fig.\,\ref{fig3}(a) right axis) and the final normalized emittance is $\epsilon_n$=180 nm. The spin vector evolutions in the $z$ directions are shown in Fig.\,\ref{fig3}(c). The spin spread in transverse ($x$ or $y$) direction is symmetric so that $\langle s_x\rangle\approx \langle s_y\rangle\approx 0$. Therefore, the net spin polarization $P=P_z=\langle s_z\rangle$ only depends on the spin distribution in the $z$ direction. The final averaged spin polarization is $\langle s_z\rangle=56\%$ [Fig.\,\ref{fig3}(c)] with almost no depolarization during injection and acceleration processes. 

\section{Discussion}
In our scheme, a good control of the delay and alignment between the driving beam and ionizing CP laser pulse is crucial, but it is possible to achieve using state-of-the-art techniques. The relative timing jitter should be controlled within tens of femtoseconds to maintain a stable beam charge and emittance. Such stringent control on the temporal jitter between a femtosecond laser pulse and the radio-frequency power source that produces the electron beam has recently been achieved in Ref. \cite{pompili2016femtosecond,kang2017hard,snively2020femtosecond,zhao2020femtosecond}. Nowadays, the angular pointing of a laser beam can be stabilized within sub-$\mu$rad level using state-of-the-art active stabilization techniques \cite{maier2020decoding, kanai2008pointing, genoud2011active, tyszka2018laser, ding2021compact}, which corresponds to sub-$\mu$m level of transverse offset fluctuation at focus with a focal length of 1 meter. We find that for the conditions of our simulation case, a transverse offset as large as $3\,\mu$m still yields the same spin polarization but the normalized emittance in the offset direction becomes twice of that in the other direction. Therefore, the pointing jitter is not thought to be a problem.

Finally, we wish to point out that the temperature dependence of the vapor pressure of Yb is very similar to that of  Li; thus it should be no more difficult to make a long homogeneous column of Yb vapor than it would be for Li which has been used with great success in the past two decades \cite{litos2014high} making the realizability of this idea promising in the near future with available facility such as FACET II. 

\section{Conclusion}
In summary, we have proposed a new scheme to produce a high-degree spin-polarized sub-femtosecond electron beam using strong-field ionization of the Yb$^{2+}$ ions by a CP laser pulse inside a plasma photocathode. Using a single atomic species, to both excite the plasma wake and be the source of spin-polarized electrons for injection, makes this concept experimentally realizable, thus solving a long-standing problem facing the development of plasma-based accelerators.

\begin{acknowledgments}
This work was supported by U.S. Department of Energy (DOE) Grant No. DE-SC0010064; DOE through a SciDAC FNAL Subcontract No. 644405; the National Science Foundation (NSF) Grants No. 1734315, No. 1806046 and No. 2108970; the Office of Naval Research (ONR) Multidisciplinary University Research Initiative (MURI) (4-442521-JC-22891); and the National Natural Science Foundation of China (NSFC) Grant No. 12075030. The simulations were performed on Hoffman cluster at UCLA and the computing resources of the National Energy Research Scientific Computing Center (NERSC).
\end{acknowledgments}

\section{Appendix}
\subsection{Details of TDSE simulations}\label{sec_TDSE}

\begin{table}[b]
\caption{\label{tab:7/2}%
Level positions for the J=7/2 core. The span of the level is the largest distance between the experimental multiplets. Ref data was from Ref.\,\cite{NIST_ASD}.
}
\begin{ruledtabular}
\begin{tabular}{cccc}
\textrm{Level}&
\textrm{Position (ref), eV}&
\textrm{Position (fit), eV}&
\textrm{Span (ref), eV}\\	
\colrule
6s & -20.738 & -20.953 & 0.041\\
7s & -10.138 & -10.220 & 0.014\\
6p & -15.570 & -15.901 & 0.886\\
7p & -7.992 & -8.269 & 0.305\\
5d & -20.092 & -19.311 & 1.270\\
6d & -9.306 & -9.220 & 0.305\\
4f & -25.053 & -25.054 & 
\end{tabular}
\end{ruledtabular}
\end{table}

\begin{table}[b]
\caption{\label{tab:5/2}%
Level positions for the J=5/2 core. The span of the level is the largest distance between the experimental multiplets. Ref data was from Ref.\,\cite{NIST_ASD}.
}
\begin{ruledtabular}
\begin{tabular}{cccc}
\textrm{Level}&
\textrm{Position (ref), eV}&
\textrm{Position (fit), eV}&
\textrm{Span (ref), eV}\\	
\colrule
6s & -20.732 & -20.926 & 0.044\\
7s & -10.138 & -10.217 & 0.011\\
6p & -15.566 & -15.911 & 0.850\\
7p & -8.216 & -8.275 & 0.303\\
5d & -20.019 & -19.397 & 1.048\\
6d & -9.516 & -9.253 & 0.494\\
4f & -26.319 & -26.318	
\end{tabular}
\end{ruledtabular}
\end{table}

The degree of spin polarization has been calculated solving the TDSE in the presence of two different single active electron potentials, using the SCID code \cite{patchkovskii2016simple}. The potentials have been modeled to describe the ionization of Yb$^{2+}$, leaving the ion in a J=7/2 or a J=5/2 state. The parameters have been fitted to the multiplet centers of mass. The summary of the reference \cite{NIST_ASD} and fitted energies relative to the $4f$ ground-state level is given in Tables\,\ref{tab:7/2} and \ref{tab:5/2}, respectively for the J=7/2 and J=5/2 cores. By adding the hard boundary to the potential, we exclude the deep $1s$, $2s$, and $2p$ levels. We furthermore do not constrain the positions of the inner levels ($3s$, $3p$, $3d$, $4s$, $4p$, $4d$, $5s$, $5p$) in the fit. These levels are never significantly populated in our TDSE simulations.

We have performed the simulations in a box of 189.17\,a.u., using a non-uniform grid, starting with a 37-point uniform grid, from 0.16\,a.u. to 1.95\,a.u. followed by a 57-point logarithmic grid with a scaling parameter of 1.025, starting at 2.0\,a.u., and ended with a 906-point uniform grid with a spacing of 0.2\,a.u from 8.17\,a.u.. In order to avoid non-physical reflections from the edges of the box, we have placed a complex absorbing potential \cite{manolopoulos2002derivation} at a distance of 156.77\,a.u. from the origin, with a width of 32.8\,a.u.. We have included up to $l$, $|m|=60$ angular channels for the angular part of the wave-function. 
We have used a right-handed CP laser pulse, with a Gaussian envelope, and a FWHM of 10 fs, with a carrier of 400\,nm (3.0996\,eV). The time coordinate was discretized with a time step of dt = 0.002 a.u.. 

Simulations were done starting from each of the possible seven $f$ initial states of Yb III ($l=3$, $m=-3$ to $m=3$), which leads to 14 simulations for each intensity point. Ionization rates as a function of intensity were calculated using the ionization probability of each channel. Summing up the ionization rates from all these different channels based on the Clebsch-Gordan coefficients \cite{kaushal2018looking1}, we can get the ionization rates of the spin-up or spin-down electrons in J=7/2 and J=5/2 states, and then get the total ionization rates of spin-up and spin-down electrons.

\subsection{Implementing TDSE ionization model into PIC code}\label{sec_TDSE_PIC}
To incorporate TDSE and PIC simulations, we implemented a new TDSE ionization model specified for photoionization of Yb$^{2+}$ ions in PIC code. This TDSE ionization model based on a series of offline TDSE simulations carried out to obtain the ionization rates of both spin-up and spin-down electrons in the range of laser intensities of interest [Fig.\,\ref{fig1}(a)]. The local photoionization rates of Yb$^{2+}$ ions in the PIC simulations are obtained via table lookup and interpolation. A certain number of macro-particles representing ionized electrons with a specific spin polarization distribution will be released according to the transient yields of spin-up  and spin-down  electrons. The density of Yb$^{2+}$ ions ($N_0$) and of Yb$^{3+}$ ions that have released spin-up ($N_\uparrow$) and spin-down electrons ($N_\downarrow$) are numerically solved through the following rate equation,
\begin{align}\label{rate_eq}
\frac{dN_0}{dt}&=-(w^\uparrow+w^\downarrow)N_0\\
\frac{dN_\uparrow}{dt}&=w^\uparrow N_0\\
\frac{dN_\downarrow}{dt}&=w^\downarrow N_0
\end{align}
where $w^{\uparrow,\downarrow}$ represents the ionization rate of spin-up or spin-down electrons.

\subsection{Details of PIC simulations}\label{sec_PIC}
The start-to-end PIC simulations consisting of two stages were carried out using the full 3D code OSIRIS \cite{fonseca2002osiris,fonseca2008one}  and the quasi-static code QPAD \cite{li2020quasi}. For each stage, the simulation window moving at the speed of light moves along the $z$ axis, i.e., the propagation direction of the driving beam and ionizing laser pulse. Ions are assumed not moving in our simulations. 

We used OSIRIS at the first stage (the injection stage) to model the photoionization, particle injection and spin precession at early times. We used ADK ionization model \cite{ammosov1986tunnel} to calculate the ionization induced by the driving electron beam and TDSE ionization model to calculate the spin-dependent photoionization rates of Yb$^{2+}$ ions induced by the ionizing laser. In our simulations, we always make sure that the driving electron beam fully ionizes the first two while not the third electron of Yb near the focus of the ionizing laser. In other words, we always make sure the ionizing laser only interacts with Yb$^{2+}$ ions. Only in this way, we can use two ionization models separately. The simulation window had a dimension of $114\times114\times153\,\mu$m in the $x$, $y$ and $z$ directions, respectively. We used $450\times450\times1200$ cells in the corresponding directions. Together with the selected time step of 0.3\,fs, the space-time resolution is sufficient to model the early-time laser photoionization, the trapping of the $4f^{14}$ electrons and their subsequent phase space evolution and spin precession. The number of macro-particles of Yb per cell was 32. The injected electrons were accelerated to ultra-relativistic energy ($\gamma\sim20$) until they were extracted and used as inputs for the second stage (the acceleration stage). 

At the second stage where the ionization injection has ceased, the trapped electron beam undergoes acceleration by the essentially non-evolving plasma wake. The quasi-static approximation \cite{sprangle1990nonlinear} is valid in the absence of particle injection and thus the quasi-static code QPAD was employed to explore the physics therein. Therefore, TDSE ionization model is not included any more, but we still use ADK ionization model to calculate the ionization induced by the driving electron beam. Benefiting from the speed-up techniques in QPAD, a long-distance (time) simulation with much finer resolution at lower cost of computational resources is achievable. The moving simulation window has a dimension of $57\times153\,\mu$m in the radial and propagation directions, with $3200\times3200$ cells in the corresponding directions. Since the selection of time step in a quasi-static code is not subject to the numerical stability consideration, a much larger time step of 78\,fs is chosen to resolve the betatron oscillation of beam particles. 


%

\end{document}